\documentclass[reprint, superscriptaddress, amsmath,amssymb, aps]{revtex4-2}

\usepackage{graphicx}%
\usepackage{xcolor}%
\usepackage{bm}
\usepackage{amsmath}

\begin{document}
\title[Observation of the anomalous Nernst effect in altermagnetic candidate $\rm Mn_5 Si_3$]{Observation of the anomalous Nernst effect in altermagnetic candidate $\rm Mn_5 Si_3$}

\author{Anton\'{i}n Badura}
\email{badura@fzu.cz}
\affiliation{%
Institute of Physics, Czech Academy of Sciences, Prague, Czechia
}%
\affiliation{%
Faculty of Mathematics and Physics, Charles University, Prague, Czechia
}%

\author{Warlley H. Campos}%
\author{Venkata K. Bharadwaj}%
\affiliation{%
Institute of Physics, Johannes Gutenberg University Mainz, Mainz, Germany
}%

\author{Isma\"{i}la Kounta}%
\author{Lisa Michez}%
\author{Matthieu Petit}%
\affiliation{%
Aix Marseille Univ, CNRS, CINAM, AMUTECH, Marseille, France
}%

\author{Javier Rial}%
\author{Miina Leivisk\"{a}}%
\author{Vincent Baltz}%
\affiliation{%
Univ. Grenoble Alpes, CNRS, CEA, Grenoble INP, IRIG-SPINTEC, Grenoble, France
}%

\author{Filip Krizek}%
\author{Dominik Kriegner}%
\affiliation{%
Institute of Physics, Czech Academy of Sciences, Prague, Czechia
}%

\author{Jan Zemen}%
\affiliation{%
Faculty of Electrical Engineering, Czech Technical University, Prague, Czechia
}%

\author{Sjoerd Telkamp}%
\affiliation{%
Solid State Physics Laboratory, ETH, Z\"urich, Switzerland
}%

\author{Sebastian Sailler}%
\author{Michaela Lammel}%
\affiliation{%
Department of Physics, University of Konstanz, Konstanz, Germany
}%

\author{Rodrigo Jaeschke Ubiergo}
\author{Anna Birk Hellenes}
\affiliation{%
Institute of Physics, Johannes Gutenberg University Mainz, Mainz, Germany
}%

\author{Rafael Gonz{\'a}lez-Hern{\'a}ndez}
\affiliation{%
Grupo de Investigación en Física Aplicada, Departamento de Física, Universidad del Norte, Barranquilla, Colombia
}%
\affiliation{%
Institute of Physics, Johannes Gutenberg University Mainz, Mainz, Germany
}%

\author{Jairo Sinova}%
\affiliation{%
Institute of Physics, Johannes Gutenberg University Mainz, Mainz, Germany
}%
\affiliation{%
Department of Physics, Texas A\&M University, College Station, Texas, USA
}%

\author{Tom\'a\v{s} Jungwirth}%
\affiliation{%
Institute of Physics, Czech Academy of Sciences, Prague, Czechia
}%
\affiliation{%
School of Physics and Astronomy, University of Nottingham, Nottingham, United Kingdom
}%

\author{Sebastian T. B. Goennenwein}%
\affiliation{%
Department of Physics, University of Konstanz, Konstanz, Germany
}%

\author{Libor \v{S}mejkal}%
\affiliation{%
Institute of Physics, Johannes Gutenberg University Mainz, Mainz, Germany
}%
\affiliation{%
Institute of Physics, Czech Academy of Sciences, Prague, Czechia
}%

\author{Helena Reichlova}%
\affiliation{%
Institute of Physics, Czech Academy of Sciences, Prague, Czechia
}%

\begin{abstract}
The anomalous Nernst effect generates transverse voltage to the applied thermal gradient in magnetically ordered systems. The effect was previously considered excluded in compensated magnetic materials with collinear ordering. However, in the recently identified class of compensated magnetic materials, dubbed altermagnets, time-reversal symmetry breaking in the electronic band structure makes the presence of the anomalous Nernst effect possible despite the collinear spin arrangement.  In this work, we investigate epitaxial $\rm Mn_5Si_3$ thin films known to be an altermagnetic candidate. We show that the material manifests a sizable anomalous Nernst coefficient despite the small net magnetization of the films. The measured magnitudes of the anomalous Nernst coefficient reach a  scale of microVolts per Kelvin. We support our magneto-thermoelectric measurements by density-functional theory calculations of the material's spin-split electronic structure, which allows for the finite Berry curvature in the reciprocal space. Furthermore, we present our calculations of the intrinsic Berry-curvature Nernst conductivity, which agree with our experimental observations.
\end{abstract}

\maketitle

\begin{figure*}%
\centering
\includegraphics{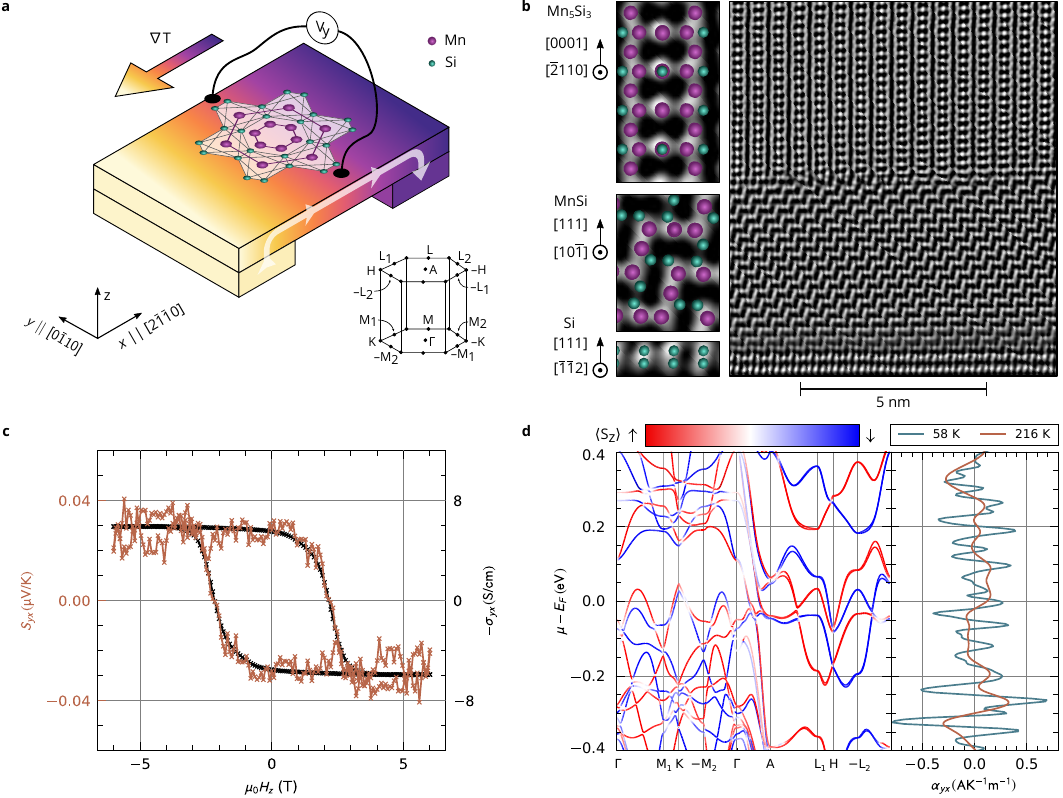}
\caption{The anomalous Nernst effect in $\rm Mn_5 Si_3$.  \textbf{a} Schematic illustration of the experiment: A longitudinal temperature gradient $\nabla T$ induces a spontaneous transverse voltage $V_{y}$.  \textbf{b} A cross-section of a typical sample, i.e. substrate / MnSi buffer layer / $\rm Mn_5 Si_3$ layer, captured by scanning transmission electron microscopy, including crystal structure models in the left panel. \textbf{c} Transverse Nernst signal $S_{yx}$ as a function of applied magnetic field for a sample temperature of 216~K. The figure also includes the field dependence of the anomalous Hall conductivity for comparison. \textbf{d} \textit{Ab initio} calculations of the band structure of $\rm Mn_5 Si_3$ (the left panel) including the calculated anomalous Nernst conductivities at 58~K and 216~K (the right panel).}\label{fig1}
\end{figure*}

The anomalous Nernst effect (ANE, see Fig. \ref{fig1}a) was traditionally considered only in ferromagnetic materials. However, recent studies have shown that it can also be observed in non-collinear antiferromagnets, such as $\rm Mn_3Sn$ \cite{ikhlas2017} and $\rm Mn_3NiN$ \cite{beckert2023anomalous}. In these materials, the ANE can arise from a finite Berry curvature in the momentum space and is allowed by the symmetry of the atomic positions and non-collinear spins of the frustrated magnetic lattice \cite{chen2014anomalous, kubler2014, nakatsuji2015, zhou2020giant}. This leads to the presence of a sizable ANE despite the vanishing net magnetization. The potential to enhance the ANE response has renewed interest in this phenomenon both in ferromagnets \cite{guin2019anomalous, chen2022large} and antiferromagnets  \cite{pan2022giant, zhou2020giant}, with possibilities for its application in heat harvesting elements or heat flux sensors. Additionally, the presence of the ANE in a material enables the use of novel magnetic imaging microscopy techniques that rely on thermal gradients \cite{weiler2012local, reichlova2019imaging, janda2020magneto, johnson2022identifying}. However, the complex spin arrangement of non-collinear antiferromagnets does not support long spin coherence, and the high ANE magnitude is often observed in noncollinear systems with heavy elements. In the broad family of compensated magnets with a robust unfrustrated collinear antiparallel ordering, the ANE was thought to be absent. In this combined experimental and theoretical work, we demonstrate that in the recently identified class of altermagnets \cite{smejkal2020crystal, smejkal2022emerging}, the spontaneous anomalous Nernst effect can be observed. Our calculations, consistent with experimental findings, show the spontaneous ANE in altermagnet candidate $\rm Mn_5Si_3$ thin films, which contain light and abundant elements. Our results suggest a new direction for observing and utilizing the ANE within the broad family of collinear compensated magnets.

In the paramagnetic state, $\rm Mn_5 Si_3$ crystallizes in a hexagonal unit cell with the space group $P6_3/mcm$. There are two non-equivalent sites of manganese atoms in its unit cell: four Mn atoms (Mn1) occupy the Wyckoff position 4d, whereas the remaining six are at the Wyckoff position 6g (Mn2) \cite{aronsson1960, gottschilch2012}. Bulk $\rm Mn_5 Si_3$ undergoes two successive first-order transitions into antiferromagnetic phases as established by neutron-diffraction experiments \cite{brown1995, gottschilch2012, biniskos2022}: Below 100~K, the crystal symmetry is reduced to orthorhombic accompanied by a collinear antiferromagnetic configuration with a symmetry combining time-reversal and half unit-cell translation. The second transition occurs at 60~K when the magnetic order becomes highly noncollinear and noncoplanar \cite{biniskos2022}. 

Due to epitaxial strain at the interface with the substrate, the in-plane lattice parameters of our epitaxial $\rm Mn_5 Si_3$ films exhibit no structural transitions, as evidenced by X-ray diffraction measurements. Consequently, the unit cell remains hexagonal in the temperature range of 10--300~K \cite{reichlova2020, kounta2023}. The expected collinear magnetic order between  $\approx 80\,\rm K$ and $\approx 240\,\rm K$ is consistent with the density-functional theory calculations and magnetotransport experiments \cite{reichlova2020}. This particular crystal and magnetic structure leads to the breaking of translation and inversion symmetries while preserving a rotation symmetry connecting sublattices with opposite spins. These non-relativistic spin symmetries categorize the material as altermagnetic \cite{smejkal2022emerging}. Experimental evidence further supports the existence of the altermagnetic phase below the ordering temperature of $\approx 240\,\rm K$ in these films by the measured anomalous Hall effect (AHE) \cite{reichlova2020, han2024electrical} despite a negligible net remanent magnetization. The AHE response in $\rm Mn_5 Si_3$ epilayers is spontaneous, i.e., it is sizable even in the absence of the magnetic field. It was ascribed to the altermagnetic mechanism of the time-reversal symmetry breaking in the electronic structure. The non-relativistic spin-split electronic structure is anisotropic in the momentum space with the d-wave symmetry \cite{reichlova2020}. Lowering the symmetry by the relativistic spin-orbit coupling then allows for a non-zero Berry curvature in the momentum space. Correspondingly, the AHE can be detected in measurements of the $(0001)$-oriented $\rm Mn_5 Si_3$ films, except for the case of the N\'{e}el vector aligned with the $[0001]$-axis for which the integral of the Berry curvature vanishes by symmetry, or for the Néel vector within the $(0001)$-plane, $(2\overline{1}\overline{1}0)$-plane or $(0\overline{1}10)$-plane for which the Hall vector, if allowed, is constrained by symmetry to the $(0001)$-plane of the thin film. The ANE, linked to the AHE by the Mott relation, obeys analogous symmetry requirements, and its existence is thus allowed.

\section*{Anomalous Nernst effect in $\rm Mn_5 Si_3$ films}
For our experiments, we use 20nm $\rm Mn_5 Si_3$(0001) films grown by molecular-beam epitaxy on intrinsic Si(111) substrates. Fig. \ref{fig1}b shows a section of our $\rm Mn_5 Si_3$ epilayer along its $[0001]$ direction as captured by scanning transmission electron microscopy (STEM). The STEM image displays the high quality of our film and its interface, as well as the buffer layer of MnSi, which forms at the interface after annealing. See Methods and Ref.  \cite{kounta2023} for more details about the layer growth and characterization. Our samples exhibit the spontaneous AHE, as shown in Fig. \ref{fig1}c (black line).

\begin{figure*}%
\centering
\includegraphics[width=\textwidth]{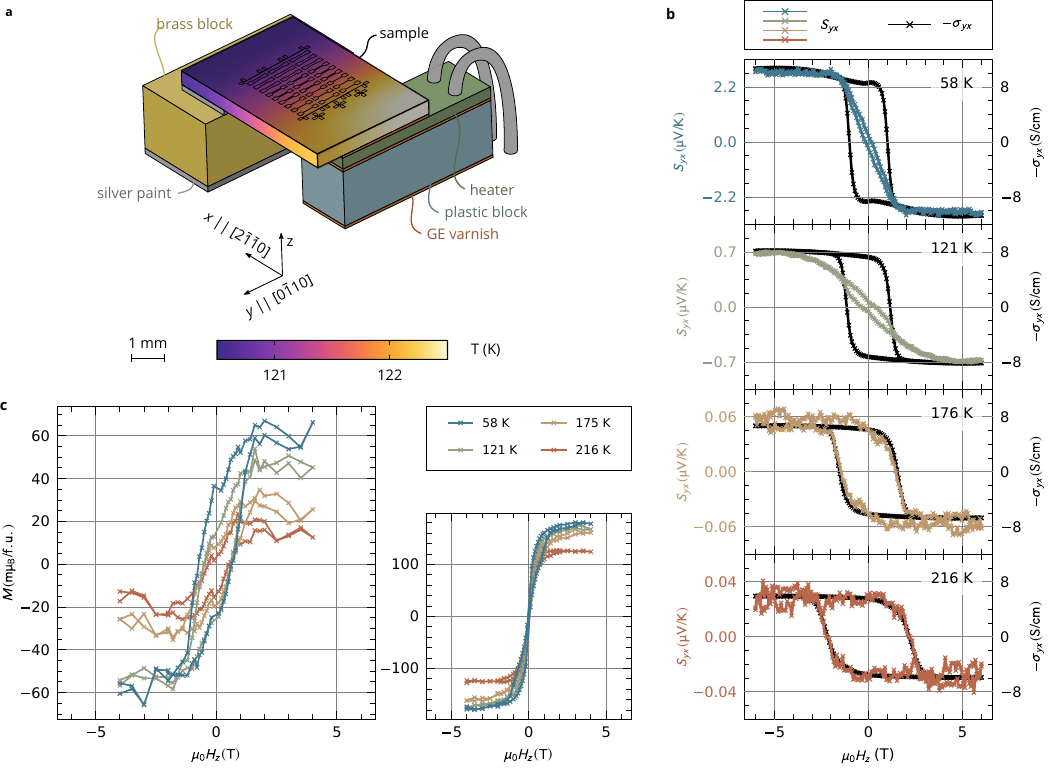}
\caption{Temperature dependence of the anomalous Nernst effect in $\rm Mn_5 Si_3$. \textbf{a} Detailed image of the experimental setup: The longitudinal temperature gradient in the sample is induced by heating up the sample with a platinum resistive heater. The colour map of the sample temperature for a particular setting was calculated by a finite-element simulation. \textbf{b} Magnetic-field dependence of the anomalous Nernst signal $S_{yx}$ for four average sample temperatures. Each figure shows the anomalous Hall conductivity measured at the same temperature for comparison. \textbf{c} Dependence of the detected out-of-plane magnetization on the applied magnetic field for different sample temperatures. The main figure features the spontaneous magnetization of our $\rm Mn_5Si_3$ layers after subtracting a saturating contribution from the substrate, whereas the raw data are given in the inset. In the presented data, the dominating diamagnetic signal of the silicon substrate has been subtracted.}\label{fig2}
\end{figure*}

To measure the ANE in our samples, we generate an in-plane temperature gradient along the $[\overline{2}110]$ direction by a macroscopic heater as illustrated in Fig. \ref{fig2}a: The sample is supported by a plastic block with a resistive heater on one side and by a brass block on the other side \cite{reichlova2018large}. The different thermal conductivities of the plastic and brass blocks enhance the temperature gradient generated by the heater. To detect the thermovoltage generated in the $\rm Mn_5 Si_3$ layer, we lithographically defined $\rm Mn_5 Si_3$ strip structures, as shown in the detail of our design in Extended Data Fig. E1. We detect the thermovoltage $V_{y}$ transverse to the gradient direction as a function of the magnetic field applied along the z-axis (film’s c-direction) as illustrated in Fig. \ref{fig1}a. An example of such a field-dependent transverse thermopower $S_{yx}$ is shown in Fig. \ref{fig1}c for a sample average temperature of 216~K. $S_{yx}$ shows clear hysteretic behaviour and saturation, a signature of the finite ANE in our samples. The saturation field and coercivity of $S_{yx}$ match the field dependence of the transverse conductivity $\sigma_{yx}$ shown in the same panel. The experimental manifestation of ANE is supported by our \emph{ab initio} calculations of the anomalous Nernst conductivity $\alpha_{yx}$. The energy dependence of $\alpha_{yx}$ for temperature 58~K and 216~K is shown in Fig. \ref{fig1}d next to the spin-resolved electronic band structure of altermagnetic $\rm Mn_5 Si_3$. It reaches the value of $0.25\,\rm A/(K\cdot m)$ at the Fermi energy for 58~K.

A crucial step in quantifying the Nernst coefficient $S_{xy}$ is to determine the temperature gradient in the sample precisely. To do so, we measure the temperature at two places on the sample by deposited platinum strips. The precise spatial temperature distribution is simulated by the finite-element method in \textsc{comsol multiphysics} \cite{comsol}, where we used realistic geometry of our experimental setup and adjusted the surface heat transfer to fit the measured sample temperatures. The result is shown as a colour map in Fig. \ref{fig2}a for the average sample temperature of  121~K. An example of the temperature gradient profile determined by the simulation is shown in Extended Data Fig. E1: The temperature gradient is linear in the central patterned part of the sample and deviates from the linearity at the edges. The magnitude of the simulated thermal gradient along the $x$-axis varied between $0.04\,\rm K/mm$ at 58~K and $0.93\,\rm K/mm$ at 216~K.

Fig. \ref{fig2}b presents $S_{yx}$ measured at four different sample temperatures between 58~K and 216~K. We take into account only the odd-in-field part of the $S_{yx}$ field dependence to remove the effect of the temperature gradient not being perfectly perpendicular to the lithographic bars. The figure also features the field dependence of the transverse conductivity $\sigma_{yx}$ measured on the same sample at the same temperatures. Both $S_{yx}$ and $\sigma_{yx}$ are presented without the linear-in-field contribution of the ordinary Nernst effect and the ordinary Hall effect, respectively. The magnitude of the saturated $S_{yx}$ is strongly temperature dependent:  It changes from $-(3\pm 2)\,\rm \mu V/K$ at 58~K to $-(0.04\pm 0.01)\,\rm \mu V/K$ at 216~K. In contrast, the saturated $\sigma_{yx}$ only slightly decreases from $(10.6\pm 0.1)\rm\, S/cm$ to $(5.91\pm 0.02)\rm\, S/cm$. We also measured the field dependence of the Seebeck coefficient $S_{xx}$. This allows us to determine the spontaneous Nernst conductivity $\alpha_{yx}$, which varies from $\alpha_{yx}=(0.11\pm0.08)\,\rm A/(K\cdot m)$ at 58~K to $\alpha_{yx}=(0.015\pm0.005)\,\rm A/(K\cdot m)$ at 216~K. 

Our $\rm Mn_5 Si_3$ samples show a sizable ANE despite their small magnetization as illustrated in Fig. \ref{fig2}c: The figure shows the magnetic-field dependence of the out-of-plane component of the magnetization measured by SQUID magnetometry at different temperatures. The main figure corresponds to the spontaneous magnetization of our $\rm Mn_5Si_3$ after subtracting the saturating contribution of the substrate and the sample holder. The raw data of the same measurements are given in the inset. layer. We present the data without the dominating diamagnetic signal of the silicon substrate. The data show a clear hysteretic behaviour with the saturation value of $(45\pm 20)\,\rm m\mu_B/f.u.$ at 58~K, which decreases to $(10\pm 8)\,\rm m\mu_B/f.u.$ at 216~K. The field dependence of the magnetization suggests that the N\'{e}el vector reversal is accompanied by small canting of the magnetic sublattice moments. The observed spontaneous moment of $(20\pm 10)\,\rm m\mu_B/f.u.$ at 58~K is of the same order as the value $40\,\rm m\mu_B/f.u.$ obtained from our \emph{ab initio} calculations.

To quantitatively understand the origin of the observed ANE, we performed first-principles calculations of the Nernst conductivity $\alpha_{yx}$. The electronic band structure (left panel in Fig. \ref{fig1}d) shows a large anisotropic spin-splitting, the hallmark of altermagnets, together with a spin-mixing induced by spin-orbit coupling. This combination allows for a finite anomalous Hall conductivity in $\rm Mn_5Si_3$ \cite{reichlova2020} and for a finite ANE. In the right panel of Fig. \ref{fig1}d, we show the calculated anomalous Nernst conductivity $\alpha_{yx}$ for different temperatures as a function of the chemical potential. The Nernst conductivity calculations yield $0.25\,\rm A/(K\cdot m)$ for 58~K and $0.12\,\rm A/(K\cdot m)$ for 216~K, which agrees with the order of magnitude determined by the experiment.

$S_{yx}$ shows a complex field-dependence compared to $\sigma_{yx}$ as illustrated in Fig. \ref{fig2}b: The anomalous Nernst and Hall dependencies are nearly identical above 170~K, while the low-temperature $S_{yx}$ have pronounced sigmoid character. This leads us to the hypothesis that the $S_{yx}$ signal is composed of at least two competing contributions with different temperature dependencies. To disentangle them, we analytically describe the low-temperature sigmoid character of the $S_{yx}$ curves by an error function. When the sigmoid contribution is subtracted, only a spontaneous part of $S_{yx}$ prevails with the identical field dependence as $\sigma_{yx}$. The analysis is robust in the whole temperature range, always giving spontaneous $S_{yx}$ proportional to $\sigma_{yx}$. This analysis strategy is illustrated in Extended Data Fig. E2. The sigmoid contribution vanishes above 170~K, as quantitatively shown in Extended Data Fig. E2c. We attribute the spontaneous component of $S_{yx}$ to be a direct consequence of the altermagnetic spin-split electronic structure of our $\rm Mn_5 Si_3$ epilayers, analogous to our previous reports of the AHE in the material \cite{reichlova2020}. The interpretation of the sigmoid contribution is less clear. It cannot be attributed to the Brillouin magnetism which would not be expected up to 100~K. Because it is present only in the lower temperature range it may be related to the complex low-temperature magnetic phase of $\rm Mn_5 Si_3$ \cite{brown1995, gottschilch2012, biniskos2022} where the topological Hall effect was reported \cite{surgers2014large, reichlova2020}. However, a systematic study of the low-temperature sigmoid contribution is beyond the scope of this manuscript. 

\begin{figure}%
\centering
\includegraphics{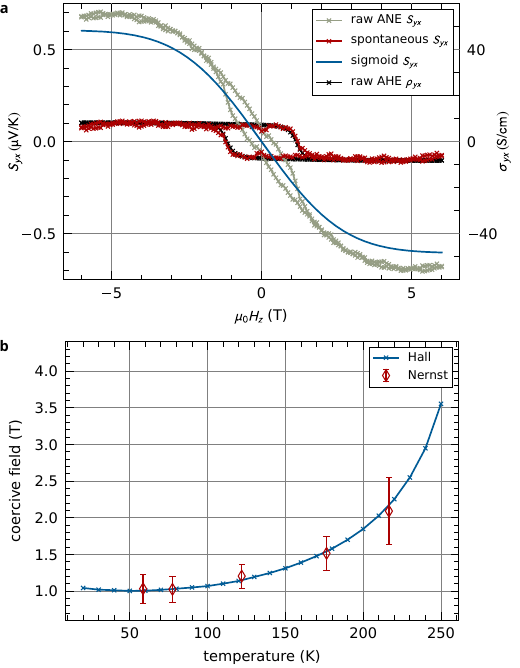}
\caption{Decomposition of the anomalous Nernst signal. \textbf{a} The anomalous Nernst signal measured at 121~K decomposed into its spontaneous and sigmoid contributions. The anomalous Hall signal is also plotted for comparison. \textbf{b} Temperature dependence of the coercive field of the spontaneous ANE contribution together with the respective dependence for the AHE.}\label{fig3}
\end{figure}

The separation of the spontaneous and sigmoid $S_{yx}$ contributions enables us to quantify the temperature dependence of the $S_{yx}$ coercive field and compare it with $\sigma_{yx}$ as shown in Fig. \ref{fig3}b. Both signals show the same increasing temperature dependence of coercivity. In a ferromagnet, the common expectation is that the coercive (reorientation) field decreases with increasing temperature, usually because of the decreasing magnetocrystalline anisotropy \cite{zener1954classical, staunton2006temperature}. The situation in compensated magnetic materials is, however, more complex: for example, the observed increase of the reorientation (spin-flop) field in collinear $\rm MnF_2$ was ascribed to its temperature-dependent and anisotropic response to the applied magnetic field \cite{shapira1969magnetic}. In our $\rm Mn_5Si_3$ thin films, however,  we do not observe any reorientation transition in the reachable magnetic field (see Fig. \ref{fig2}c). Instead, we expect that the hysteresis in the N\'{e}el vector reversal originates from the competition of the magnetocrystalline anisotropy and the Dzyaloshinskii-Moriya interaction. These effects have generally different temperature dependencies, which can result in an increasing coercivity with temperature. The temperature dependence of the magnetocrystalline anisotropy may be affected by the complex temperature dependence of the lattice parameters \cite{reichlova2020}.

\begin{figure*}%
\centering
\includegraphics[width=\textwidth]{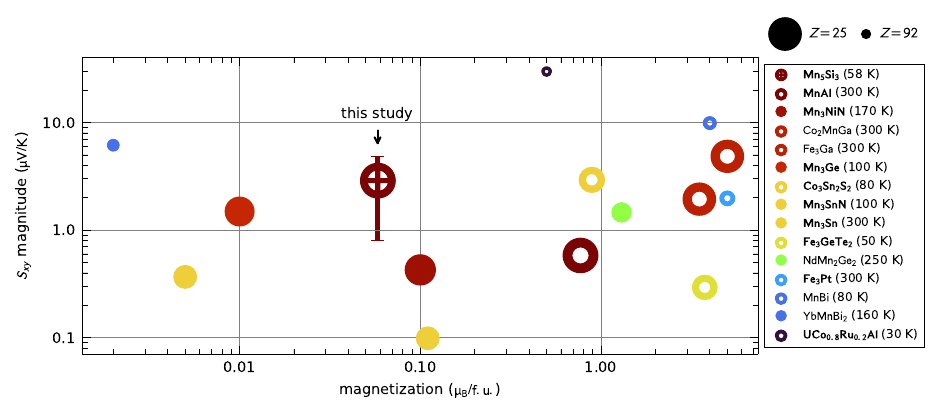}
\caption{Comparison of the anomalous Nernst effect measured in epitaxial $\rm Mn_5 Si_3$ and in other materials \cite{pan2022giant, sakai2020iron, ikhlas2017, guin2019anomalous, asaba2021colossal, he2020magnon, wuttke2019berry, beckert2023anomalous, li2023large, scheffler2023anomalous, xu2023large, reichlova2018large, you2022anomalous, xu2019large}. The ANE magnitude $S_{xy}$ is captured as a function of materials' magnetization. The symbol size linearly decreases with the highest atomic number of materials' formula. Antiferromagnetic materials are indicated by full circles, while ferromagnets are denoted by hollow circles and the altermagnet by a crossed circle. Materials showing the spontaneous Nernst effect are labelled in bold font.
}\label{fig4}
\end{figure*}

\section*{Discussion}
We have demonstrated a significant and spontaneous ANE in an altermagnetic candidate composed of light elements. Fig. \ref{fig4} presents a comparison of the ANE coefficients observed in various magnetic compounds depending on their magnetization per formula unit, including ferromagnets (hollow circles), non-collinear antiferromagnets (full circles), and our altermagnetic $\rm Mn_5Si_3$ (crossed circle). Our material showcases ANE values comparable to those of other materials, despite having vanishing magnetization, collinear magnetic order, and being constituted of light elements. 

Although we observe finite but small saturation magnetization of $45\,\rm m\mu_B/f.u.$, let us emphasise that it is not the prime source of  AHE and ANE as shown by the first-principles calculations. Instead, the AHE and ANE are dominated by the strong altermagnetic breaking of the time-reversal symmetry in the electronic structure in the absence of a net magnetization, combined with the relativistic spin-orbit coupling. An absence of a correlation between the atomic mass and the ANE is highlighted in Fig. \ref{fig4} where the symbols' size linearly decreases with the highest atomic number of the compound (the largest symbols represent the lightest elements). This observation highlights the importance of not limiting the search for high ANE performance materials solely to heavy elements. Additionally, we note that, unlike many materials listed in Fig. \ref{fig4}, our $\rm Mn_5Si_3$ is in thin film form, which often reduces the ANE magnitude. Therefore, we anticipate that higher ANE values could be discovered in altermagnetic bulks.

\section*{Conclusion}
In summary, we provide experimental evidence that altermagnetic materials can host a spontaneous ANE despite their collinear magnetic order and vanishing net magnetization, and without the need of heavy elements. We demonstrate a sizable anomalous Nernst coefficient in the altermagnetic candidate $\rm Mn_5Si_3$ in the form of epitaxial thin layers. The effect reaches $S_{yx}=-(3\pm 2)\,\rm \mu V/K$ and spontaneous $\alpha_{yx}=(0.11\pm 0.08)\,\rm A/(K\cdot m)$ at 58~K, and drops to $S_{yx}=-(0.04\pm 0.01)\,\rm \mu V/K$  and spontaneous $\alpha_{yx}=(0.015\pm 0.005)\,\rm A/(K\cdot m)$ at 216~K. The Nernst effect measurements are supported by our first-principles calculations of the $\rm Mn_5Si_3$ electronic band structure and the Nernst conductivity, which yield $\alpha_{yx}=0.25\,\rm A/(K\cdot m)$ at 58~K.

By experimentally demonstrating the ANE in the collinear altermagnets, we introduce a new category of compensated magnetic materials for exploring the ANE. This expands the field beyond the more complex non-collinear magnetic structures or materials with heavy elements, marking a significant broadening of the scope for the ANE research. The search for materials with high and robust ANE magnitude is one of the key challenges for fully exploiting their application potential. 

\section*{Methods}
\paragraph{Sample preparation} 
The $\rm Mn_5 Si_3$ thin films were grown by molecular beam epitaxy (MBE) in an ultra-high vacuum system with a base pressure better than $2\cdot 10^{-7}\,\rm Pa$.  Elemental Mn and Si were flux matched in a $5:3$ ratio using a quartz microbalance and co-deposited on the intrinsic Si(111) substrate ($R>10000\,\rm \Omega\cdot cm$) heated to $170\rm\,^\circ C$. The films were subsequently annealed up to $\approx 300\rm\,^\circ C$, the temperature for which a clear $(\sqrt{3}\times\sqrt{3})R\,30^\circ$ reconstruction is observed on the RHEED pattern. Ex situ X-ray diffraction (XRD) using a high brilliance rotating anode showed a good crystalline quality of the $\rm Mn_5 Si_3$(0001) thin films whose growth was promoted by the formation of a thin MnSi(111). A more detailed description of the growth process can be found in \cite{kounta2023}.

For the measurement of the anomalous Nernst effect, we used optical lithography and argon ion beam etching to pattern our $\rm Mn_5Si_3$ layers into two distinct lithographic designs illustrated in the Extended Data Fig. E1. The platinum thermometers of $\approx 50\,\rm nm$ thickness were deposited by magnetron sputtering. 

\paragraph{Magnetometry measurements}
Magnetic characterization of our $\rm Mn_5Si_3$ thin films was carried out using a SQUID MPMS XL 5.0 magnetometer. The as-grown sample was cleaned with acetone, alcohol, and DI water, before measurement and mounted with two dots of glue (X60) on a sample holder made of two quartz capillaries. Field-dependent magnetization was measured at different temperatures for magnetic field strengths ranging from $\pm 4\rm\, T$ applied along the film normal. The signal is dominated by the diamagnetism of the silicon substrate, which accounts for 96~\% of the signal at 4~T. After eliminating the linear-in-field diamagnetic contribution, the magnetic response contains two components: a non-hysteretic part linked to the substrate and sample holder and a hysteretic part that we attribute to the $\rm Mn_5Si_3$ film, due to a canting of the moments of the two sublattices.

\paragraph{High-resolution scanning transmission electron microscopy imaging and lamellae preparation}
The cross-sectional lamellae of selected samples were prepared by Focused Ion Beam (FIB) (Helios 5 UX from Thermo Fisher Scientific) using AutoTEM 5 software (Thermo Scientific, the Netherlands) at ScopeM, ETH Zurich. A protective carbon layer was deposited on the selected region of interest first by an electron beam (2~KV, 13~nA) and subsequently by an ion beam (30~kV, 1.2~nA). The chuck milling and lamellae thinning were done at 30~kV with FIB current from 9~nA to 90~pA. Finally, the lamellae were polished at 5~KV (17~pA) and finished at 2~kV (12~pA). The expected thickness was below 50 nm. The samples were kept in vacuum between the FIB and TEM measurements (except for loading and unloading).

The STEM images were acquired with a JEOL ARM200F cold field emission gun scanning transmission electron microscope located at the Binnig and Rohrer Nanotechnology Center Noise-free laboratories at IBM Research Europe. The images were acquired at 200 kV and were further processed using FIJI - Fiji is just ImageJ freeware \cite{schindelin2012fiji}. The FFT pattern of the images was masked by a circular aperture. After masking, a constant threshold was subtracted before performing an inverse FFT of the spectra to remove the background between the high-intensity spots corresponding to
Bragg peaks. In selected images, the contrast was normalized by the contrast normalization plugin for better visualisation of lighter elements.

\paragraph{Density-functional theory calculations}
We calculated the electronic structure of Mn$_5$Si$_3$  (space group $P6_3/mcm$, No. 193) in the pseudo-potential DFT code Vienna Ab initio Simulation Package (VASP) \cite{kresse1996efficient} within Perdew–Burke-Ernzerhof (PBE) + SOC generalized gradient approximation (GGA) \cite{kresse1996efficiency, blochl1994projector, Perdew1996generalized}. The lattice constants for the hexagonal unit cell containing two formula units (f.u.) are $a=b=6.901\,${\AA} and $c=4.795\,${\AA} -- consistent with the experimental values reported in Refs. \cite{kounta2023, reichlova2020}. We switched off symmetrization and converged the self-consistent calculations on a $9\times 9\times 12$ k-point grid using an energy cutoff of 520~eV, energy convergence criterion of $0.5\times 10^{-8}\,$eV and the tetrahedron smearing method with Blöchl corrections and smearing parameter 0.1~eV. The converged ground state has Fermi energy 9.945~eV and a total magnetization of 40~m$\mu_B$/f.u., where $\mu_B$ is the Bohr magneton. We plotted the spin-resolved electronic band structure (left panel in Fig. \ref{fig1}d) using the PyProcar Python library for electronic structure pre/post-processing \cite{Herath2020pyprocar}.

To calculate the anomalous Nernst conductivity (ANC; right panel in Fig. \ref{fig1}d) from our DFT results, we initially obtained the Wannier functions of Mn$_5$Si$_3$ using the Wannier90 code \cite{mostofi2014updated}. Then, we incorporated these Wannier functions on the WannierBerri python library \cite{tsirkin2021high} using a $ 26 \times 26 \times 30 $ k-point grid and a Fermi-Dirac smoother with smearing parameter $0.86$~meV
to evaluate the quantities reliant on the Berry curvature \cite{Nagaosa2010anomalous}%
\begin{multline}
        \Omega^n_{yx}(\mathbf{k}) = - i \left[ \frac{\partial}{\partial {k_x}} \left\langle u_{n\mathbf{k}}\middle|\frac{\partial}{\partial {k_y}} \middle| u_{n\mathbf{k}} \right\rangle \right. \\ \left.- \frac{\partial}{\partial {k_y}} \left\langle u_{n\mathbf{k}}\middle|\frac{\partial}{\partial {k_x}} \middle| u_{n\mathbf{k}} \right\rangle \right],
\end{multline}
where $\mathbf{k}=(k_x,k_y,k_z)$ is the crystal momentum and $| u_{n\mathbf{k}} \rangle$ is the periodic part of the Bloch state with band index $n$.  Finally, we obtained the ANC for temperature $T=1$~K via the Mott relation valid in the low-temperature limit \cite{xiao2006berry}%
\begin{equation}\label{eq:ANC_lt}
    \alpha_{yx}(\mu) \approx \frac{\pi^2}{3}\frac{k_B^2 T}{e}\sum_n\int_{BZ}\frac{d\mathbf{k}}{(2\pi)^3}\Omega^n_{yx}(\mathbf{k})\frac{\partial f(\varepsilon_{n\mathbf{k}},\mu)}{\partial \varepsilon_{n\mathbf{k}}},
\end{equation}%
where $e$ is the electron elementary charge, $\mu$ is the chemical potential, $\varepsilon_{n\mathbf{k}}$ are the energy eigenvalues and $f(\varepsilon,\mu)=\left\{1+\exp{\left[\left(\varepsilon-\mu\right)/k_BT\right]}\right\}^{-1}$ is the Fermi-Dirac distribution with $k_B$ the Boltzmann constant. The integration runs over the whole Brillouin zone (BZ).

For the ANC at higher temperatures, we first obtained the anomalous Hall conductivity (AHC) at $T=0$~K \cite{Nagaosa2010anomalous}%
\begin{equation}
        \sigma_{yx}^{T=0}(\varepsilon) = -\frac{e^2}{\hbar}\sum_n\int_{BZ}\frac{d\mathbf{k}}{(2\pi)^3}\Omega^n_{yx}(\mathbf{k})\Theta (\varepsilon-\varepsilon_{n\mathbf{k}}),
\end{equation}%
where $\Theta (x)$ is the Heaviside function. Then, we used our AHC results - consistent with the calculations previously reported in Ref. \cite{reichlova2020} - to obtain the ANC via the Mott relation \cite{zhou2024crystal, xiao2006berry, ashcroft2022solid, van1992thermo, behnia2015fundamentals}
\begin{equation}\label{eq:ANC_ht}
    \alpha_{yx}(\mu)=-\frac{1}{e}\int d\varepsilon\frac{\partial f (\varepsilon,\mu)}{\partial \mu}\sigma_{yx}^{T=0}(\varepsilon)\frac{\varepsilon-\mu}{T}.
\end{equation}%

At low temperatures, Eq. (\ref{eq:ANC_ht}) reduces to $\alpha_{yx}\approx \frac{\pi^2}{3}\frac{k_B^2 T}{e}\frac{d\sigma_{yx}}{d\varepsilon}$, which we used to verify our results for $T=1$~K obtained from Eq. (\ref{eq:ANC_lt}).

We test our Nernst conductivity calculations on iron. For iron at 300~K, we obtain $\alpha_{xy}=-0.31\,\rm A/(K\cdot m)$ which is in very good agreement with the experimental observation of $\alpha_{xy}\approx -0.5\,\rm A/(K\cdot m)$ \cite{watzman2016magnon}.

\paragraph{Magnetotransport measurements}
The transport experiments were conducted in an Oxford Instruments Integra AC cryostat equipped with a 3D superconducting magnet. The temperature gradient in the sample was generated by an external heater as illustrated in Fig. \ref{fig2}a. As a heater, we used the platinum resistor Pt2000. The applied heating power ranged from 0.2 to 2~W. The sample temperature was detected by two on-chip platinum strips located on the opposite ends of the sample with respect to the applied in-plane temperature gradient. We calibrated the resistance of the platinum thermometers by a slow cool-down of the sample. The transverse Nernst voltage was detected by Keithley Nanovoltmeters 2182A. The Nernst coefficient $S_{yx}$ was determined from the measured voltage $V_{y}$ and the simulated temperature gradient $x$-component $(\nabla T)_x$ as: $S_{yx}=V_{y}/(l\cdot(\nabla T)_x)$ where $l$ is the length of the transverse $\rm Mn_5 Si_3$ contact. The main contribution to the error of the $S_{yx}$ is the determination of the temperature gradient, which we determine with the absolute error of $0.02\,\rm K/mm$. 

To evaluate the magnetic field dependence of $S_{yx}$, we only consider its odd-in-field component and subtract the linear-in-field contribution of the ordinary Nernst effect. An example of the data processing, including the raw data, is shown in Extended Data Fig. E2a. We suppose that the strong even-in-field contribution corresponds to the magneto-Seebeck effect arising due to the misalignment of contact bonds.

We determined the Nernst conductivity $\alpha_{yx}$ via the relation 
\begin{equation}
    \alpha_{yx} = \sigma_{xx}\cdot S_{yx} + \sigma_{yx} \cdot S_{xx},
\end{equation}
where $S_{xx}$ is the Seebeck coefficient and $\sigma_{xx}$ is longitudinal conductivity \cite{behnia2015fundamentals}. The latter two quantities were measured at a separate sample with a lithographically patterned Hall bar (see Extended Data Fig. E1b). $S_{xx}$ was determined in the same experimental geometry as $S_{xy}$, i.e. in the elevated configuration illustrated in Fig. \ref{fig2}a. The measured field-dependencies of $S_{xx}$ and $\sigma_{xx}$ are shown in Extended Data Fig. E3a. We present the $\alpha_{yx}$ field dependencies at different temperatures in Extended Data Fig. E3b.

\paragraph{Finite-element simulations of temperature distribution}
We simulated the distribution of the temperature in our sample by the finite-element method (FEM) in \textsc{comsol multiphysics} \cite{comsol}. We used realistic simulation geometry illustrated in Fig. \ref{fig2}a, which precisely reflects our experimental setup, including the chip carrier, the acrylonitrile styrene acrylate (ASA)
 and brass blocks, the platinum heater (in an alumina ceramic package), as well as the sample itself. The bonding wires were also taken into account. The simulation was performed using material-specific physical parameters as illustrated in Tab. \ref{tab:parameters} showing the used thermal conductivity and heat capacity values. For certain materials, we used temperature-dependent material properties.
\begin{table}[h]
\begin{tabular}{l r r} 
 \hline
  \hline
material & $C_p\,\rm (J\, kg^{-1} K^{-1})$ & $\kappa\rm\,(W\, K^{-1}m^{-1})$ \\ [0.5ex] 
 \hline
silicon & *\cite{flubacher1959heat} & *\cite{glassbrenner1964thermal}\\
brass & 380& 40 \\
ASA & 1300& 0.17\\
GE varnish & 983 & *\cite{tsatis1987thermal} \\
aluminium & 900 & 238\\
alumina ceramics & 730 & 23\\
 \hline
  \hline
\end{tabular}
\caption{Values of heat capacity at constant pressure $C_p$ and thermal conductivity $\kappa$ used for the finite-element modelling of the temperature gradient. For values marked by an asterisk, we used temperature-dependent $C_p$ or $\kappa$.}
\label{tab:parameters}
\end{table}

Our FEM model includes the heat equation with the mixed boundary condition of the constant temperature at the interfaces in contact with the chip carrier and constant heat flux elsewhere. The heat transfer coefficient is a fitting parameter of our model that we adjust to reach the temperatures experimentally observed on the on-chip platinum thermometers. The coefficient is temperature-dependent and increases from $26\,\rm W/(m^2\cdot K)$ at 58~K to $89\,\rm W/(m^2\cdot K)$ at 216~K. The sample heating is simulated via the Joule heating by the charge current in the platinum heater. 

As illustrated in the Extended Data Fig. 1a, the simulated temperature gradient is linear in the measurement area of the sample, where it reaches $-\partial T/\partial x = 0.04\,\rm K/mm$ for the sample temperature of 58~K, $0.34\,\rm K/mm$ for 121~K, $0.74\,\rm K/mm$ for 176~K, $0.93\,\rm K/mm$ 216~K. Outside the measurement area, the temperature gradient is not monotonous and gains a finite out-of-plane component.

\section*{Acknowledgements}
The authors acknowledge the support of Peng Zheng from ScopeM ETH Zurich and the support of Marilyne Sousa (IBM) via the Binning and Rohrer Nanotechnology Center (BRNC). The study was supported by Charles University, project GA UK No. 266723, by the Deutsche Forschungsgemeinschaft (DFG, German Research Foundation) (Project-IDs 445976410 and 490730630), by the French National Research Agency (ANR) (Project MATHEEIAS - Grant No. ANR-20-CE92-0049-01 and Project PEPR SPIN – Grant No. ANR-22-EXSP-0007), and by the Grant Agency of the Czech Republic Grant No. 22-17899K and the Max Planck Dioscuri Program. D.K. acknowledges the support from the Czech Academy of Sciences (project No. LQ100102201). 
\bibliography{bibliography}
\end{document}